\documentclass[sn-aps]{sn-jnl}

\jyear{2023}%

\begin{document}

\title[]{Quantum Approximate Optimization via Learning-based Adaptive Optimization}


\author*[1]{\fnm{Lixue} \sur{Cheng}}\email{sherrycheng@tencent.com}
\equalcont{These authors contributed equally to this work.}

\author[1]{\fnm{Yu-Qin} \sur{Chen}}
\equalcont{These authors contributed equally to this work.}

\author*[1]{\fnm{Shi-Xin} \sur{Zhang}}\email{shixinzhang@tencent.com}


\author*[1]{\fnm{Shengyu} \sur{Zhang}}\email{shengyzhang@tencent.com}

\affil[1]{\orgname{Tencent Quantum Laboratory}, \orgaddress{\city{Shenzhen}, \postcode{518057},\country{China}}}


\maketitle

\begin{abstract}

\section*{Abstract}

\noindent{Combinatorial optimization problems are ubiquitous and computationally hard to solve in general. 
Quantum approximate optimization algorithm (QAOA), one of the most representative quantum-classical hybrid algorithms, is designed to solve combinatorial optimization problems by transforming the discrete optimization problem into a classical optimization problem over continuous circuit parameters. QAOA objective landscape is notorious for pervasive local minima, and its viability significantly relies on the efficacy of the classical optimizer. 
In this work, we design double adaptive-region Bayesian optimization (DARBO) for QAOA. Our numerical results demonstrate that the algorithm greatly outperforms conventional optimizers in terms of speed, accuracy, and stability. We also address the issues of measurement efficiency and the suppression of quantum noise by conducting the full optimization loop on a superconducting quantum processor as a proof of concept. This work helps to unlock the full power of QAOA and paves the way toward achieving quantum advantage in practical classical tasks.}
\end{abstract}


\maketitle

\section{Introduction}\label{sec:introduction}
Combinatorial optimization, which involves identifying an optimal solution from a finite set of candidates, has a wide range of applications across various fields, such as logistics, finance, physics, and machine learning. However, the problem in many typical scenarios is NP-hard since the set of feasible solutions is discrete and expands exponentially with the growing problem size without any structure that seems to admit polynomial-time algorithms. 
As a representative NP-hard problem, MAX-CUT aims to find a bi-partition of the input graph's vertices, such that the number of edges (or total edge weights) between the two subsets reaches the maximum. 
Classical approaches such as greedy algorithms and AI methods by graph neural networks, despite remarkable attempts and progresses \cite{Schuetz2022, Angelini2023, Boettcher2023}, are generally inefficient to address combinatorial optimization problems such as MAX-CUT due to their NP-hard nature. In the recent two decades, quantum computing approaches have emerged as a new toolbox for tackling these difficult but crucial problems, including quantum annealing \cite{Kadowaki1998, Farhi2001, Johnson2011, Hauke2020, Hibat-Allah2021} and quantum approximate optimization algorithm (QAOA) \cite{Farhi2014, zhou2020quantum, Arute2020, Larkin2022, Pelofske2023}, from both theoretical and experimental perspectives. In this article, we focus on the latter paradigm, which is fully compatible with the universal gate-based quantum circuit model and is considered to be one of the most promising algorithms in the noisy intermediate-scale quantum (NISQ) era for potential quantum advantages.

In the QAOA paradigm, the exponential solution space is encoded in the Hilbert space of the output wavefunction of a parameterized quantum circuit. By this, the classical optimization problems in the discrete domain are relaxed to a continuous domain composed of circuit variational parameters via QAOA as a proxy. However, the classical optimization over the continuous circuit parameter domain is still challenging (the worst case is NP-hard \cite{Bittel2021}) since the energy landscape of the QAOA ansatz is filled with local minima and a large amount of independent optimization processes is required to identify a near-optimal solution \cite{Anschuetz2021,zhou2020quantum}. In addition, barren plateaus can also emerge in the QAOA landscape with increasing qubit number or circuit depth \cite{McClean2018, Marrero2020, Wang2020, Arrasmith2020}. To overcome these optimization difficulties, various learning-based \cite{Verdon2019a, Alam2020, Khairy2020, Jain2022, Shaydulin2021, Moussa2022, Amosy2022, Yao2022,Xie2022} or heuristic-based approaches \cite{zhou2020quantum,Tate2022, Campos2021, Shaydulin2022, Sack2022, mele2022avoiding} have previously been explored. These methods either rely on optimization data previously obtained or require a huge number of circuit evaluation budgets in total by progressively searching solutions of QAOA with different depths. A universally efficient and effective optimization approach suitable for real quantum processors without prior knowledge remains elusive.

In this work, we design a gradient-free classical optimizer dubbed Double Adaptive-Region Bayesian Optimization (DARBO), which exploits and explores the QAOA landscape with a Gaussian process (GP) surrogate model and iteratively suggests the most possible optimized parameter set restricted by two auto-adaptive regions, i.e., an adaptive trust region and an adaptive search region.
The performance of DARBO for QAOA and ultimately for combinatorial optimization problems in terms of speed, stability, and accuracy is superior to existing methods. Furthermore, DARBO is robust against measurement shot noise and quantum noise. We demonstrated its effectiveness in extensive numerical simulations as well as a proof of concept demonstration of the quantum-classical optimization pipeline, where QAOA is implemented and evaluated on a real superconducting quantum processor using five qubits with integrated quantum error mitigation (QEM) techniques.

\begin{figure*}[htp]
    \centering    \includegraphics[width=1\textwidth]{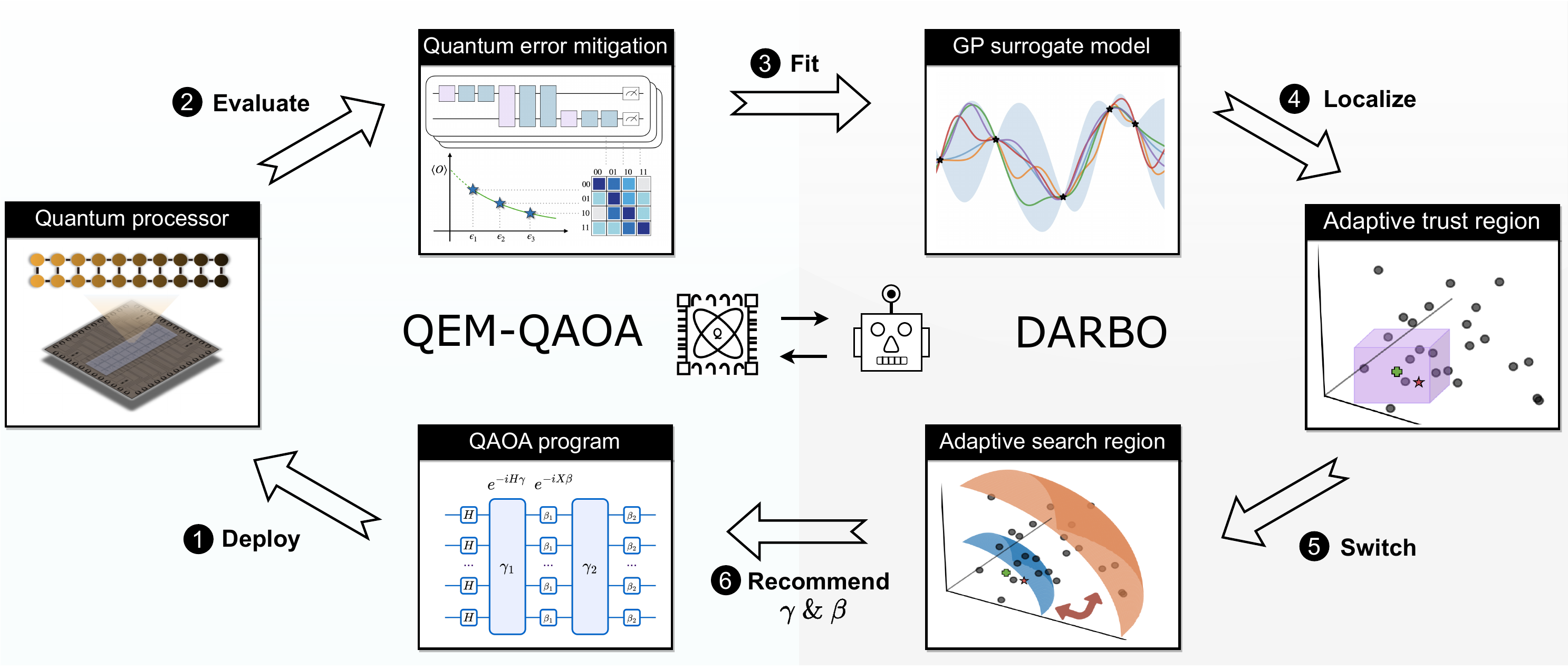}
    \caption{The proof-of-concept workflow for error mitigated QAOA on the superconducting quantum processor with DARBO. We compile and deploy the 5-qubit QAOA program for given objective functions on a 20-qubit real superconducting quantum processor and evaluate the objective value with quantum error mitigation methods. DARBO treats the QEM-QAOA as a black-box, and optimizes the circuit parameters by fitting the surrogate model with constraints. The constraints are provided by the two adaptive regions, which are responsible for surrogate model building and acquisition function sampling, respectively.}
    \label{fig:fig1}
\end{figure*}

\begin{figure*}[t]
    \centering
    \includegraphics[width=1.0\textwidth]{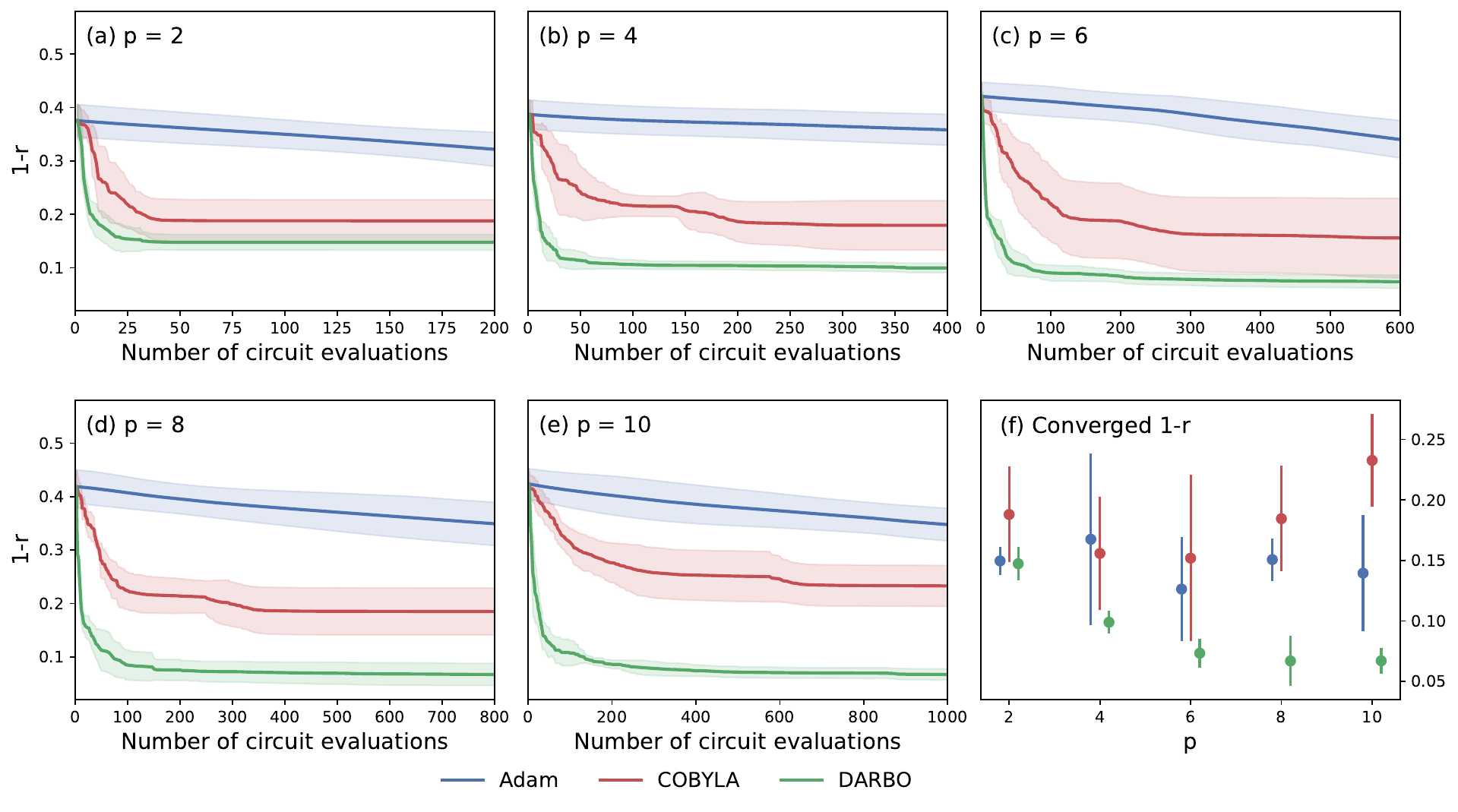}
    \caption{QAOA optimization for MAX-CUT problem on w3R graphs (exact simulations). All the optimizations are performed on $n=16$ w3R graphs. (a)-(e) The optimization trajectories from different optimizers, i.e., Adam (in blue color), COBYLA (in red color), and DARBO (in green color) are plotted versus the number of circuit evaluations. Results on different circuit depths from $p=2$ to $p=10$ are reported, respectively. (f) The final converged approximation gap $1-r$ after sufficient numbers of optimization iterations. For each circuit evaluation, we query the exact expectation of the objective function via numerical simulation. Each line is averaged over five w3R graph instances, where the shaded range shows the standard deviation of the results across different graph instances. For each graph instance, the best optimization result among the 20 independent optimization trials is reported. The error bar in (f) shows the standard deviation across different graph instances. }
    \label{fig:fig2}
\end{figure*}

\section{Results}\label{sec:results}
\subsection{QAOA framework with DARBO}\label{sec:qaoa}

MAX-CUT problem and a large family of combinatorial optimization problems can be embedded into a more general formalism as quadratic unconstrained binary optimization (QUBO) \cite{Norouzi2009}, where the objective function $C(\bf{z})$ to optimize is in the form as follows in terms of binary-valued variables $\bf{z}$:
\begin{equation}
{C(\mathbf{z}) = \sum_{ij}w_{ij}z_iz_j,}
\end{equation}
where $w_{ij}$ can be regarded as the edge weights defined on a given graph.

The QAOA framework is designed as a quantum-enhanced method to solve these QUBO problems. The quantum circuit ansatz for QAOA consists of the repetitive applications of two parameterized unitary operators. We denote the number of repetitions as $p$, and the number of qubits (binary freedoms in QUBO) as $n$.  The quantum program ansatz is constructed as 
\begin{align}
&\vert \psi(\mathbf{\gamma}, \mathbf{\beta})\rangle=
U(\mathbf{\gamma}, \mathbf{\beta})\vert 0^n\rangle = \nonumber\\ &\prod_{k=1}^p \left(e^{-i\beta_k \sum_i^n X_i} e^{-i\gamma_k \sum_{ij}w_{ij} Z_iZ_j} \right)\prod_i^n H_i\vert 0^n\rangle,
\end{align}
where $X$ and $Z$ are Pauli matrice on each site and $H$ are Hadamard gate. 
The trainable parameters $\bf{\gamma}$ and $\bf{\beta}$ both consist $p$ real-valued components, and the outer classical training loop adjusts these parameters so that the objective $C(\gamma, \beta)=\langle \psi(\gamma, \beta)\vert \sum_{ij}w_{ij}Z_iZ_j \vert \psi(\gamma, \beta)\rangle$ is minimized. Therefore, by utilizing the QAOA framework, the optimization over discrete binary $\bf{z}$ variables is reduced to the optimization over continuous variables of $\beta$ and $\gamma$.

However, the continuous optimization problem still faces lots of pressing challenges. In the QAOA framework, gradient descent optimizers commonly utilized in the deep learning community do not work well. 
The common ansatze consisting of a large number of parameters can enjoy the benefits of over-parameterization and their global minima are easier to locate \cite{larocca2021theory, Kim2021}.  However, QAOA ansatz has a small number of parameters and thus a large number of local minima, which often destroys the effort of conventional optimizers to identify the global minimum.
Besides, barren plateaus that render the gradient variance vanishing exponentially can occur similarly as the generic cases in variational quantum algorithms. More importantly, gradient evaluations on real quantum chips are too noisy and costly to use for a classical optimizer. 

Bayesian optimization (BO) is a class of black-box and gradient-free classical optimization approaches that can effectively optimize expensive black-box functions and tolerate stochastic noise in function evaluations. The method typically creates a surrogate for the unknown objective, and quantifies and manages the uncertainty using a Bayesian learning framework \cite{frazier2018tutorial,eriksson2019scalable}. Although conventional BO has become a highly competitive technique for solving optimizing problems with a small number of parameters, it usually does not scale well to problems with high dimensions \cite{frazier2018tutorial,eriksson2019scalable,Letham2017ConstrainedBO,letham2020re}. Aside from the plentiful local minima in the exponentially large search space, another challenge with BO is that the surrogate function fitting with very few samples can hardly be globally accurate. 

To overcome the above issues and enable efficient QAOA executions on real quantum chips, we propose DARBO as a powerful classical optimizer for QAOA. The schematic quantum-classical hybrid workflow for DARBO-enabled QAOA is shown in Fig.~\ref{fig:fig1}. The advantages of DARBO are both from its Bayesian optimization nature and the two adaptive regions utilized in the algorithm. The idea of including an adaptive trust region is directly borrowed from TuRBO \cite{eriksson2019scalable} and is inspired by a class of trust region methods from stochastic optimization \cite{yuan2015recent}. These methods utilize a simple surrogate model inside a trust region centered on the current best solution. For instance, COBYLA \cite{Powell1994} method used as a baseline in this work models the objective function locally with a linear model. As a deterministic approach, COBYLA is not good at handling noisy observations. By integrating with GP surrogate models within an adaptive trust region, DARBO inherits the robustness to noise and rigorous reasoning about uncertainties that global BO enjoys as well as the benefits that local surrogate model enables.
In addition, the introduction of an adaptive search region makes DARBO more robust to different initial guesses by moving queries in some iterations to a more restricted region. The search efficiency increases when the search space is reduced by the adaptive search region, giving DARBO a higher chance of finding the global minimum rather than local minima.

In this study, the end-to-end performance of QAOA with DARBO is evaluated and benchmarked on the basis of analytical exact simulation, numerical simulation with measurement shot noise, and quantum hardware experiments (with both measurement shot noise and quantum noise). Overall, DARBO outperforms other common optimizers by a large margin in terms of 1) efficiency: the number of circuit evaluations to reach a given accuracy is the least, 2) stability: the fluctuation of the converged objective value across different initializations and graph instances is the least, 3) accuracy: the final converged approximation ratio is the best, and 4) noise robustness: the performance advantage is getting larger when noise presents, which is unavoidably the case on quantum processors.

We use approximation ratio $r$ as a metric to measure the end-to-end performance of QAOA. In MAX-CUT problem, $r$ is defined as the ratio of obtained cut value derived from the objective expectation over the exact max cut value ($r=1$ indicates a perfect solution for the problem), and $1-r$ is the (relative) approximation gap. Throughout this work, we investigate the MAX-CUT problem on the w3R graph which is a family of regular weighted graphs whose vertices all have degrees of three.

\hspace*{\fill}

\begin{figure*}[tp]
    \centering    \includegraphics[width=1.0\textwidth]{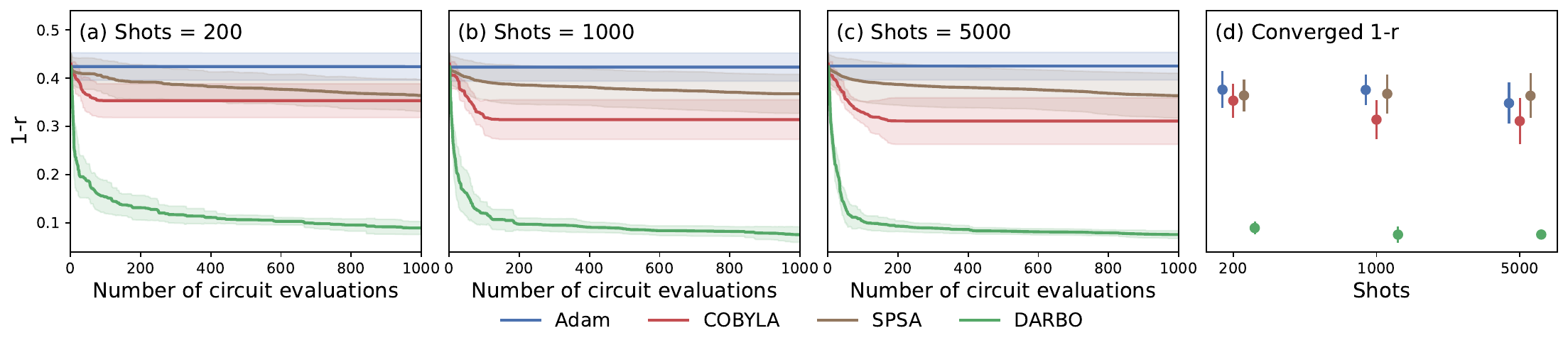}
    \caption{QAOA optimization for MAX-CUT problem on w3R graph (simulation with measurement shot noise). All the optimizations are performed on $n=16$ w3R graphs with $p=10$. (a)-(c) The optimization trajectories from different optimizers, i.e., Adam (in blue color), COBYLA (in red color), SPSA (in brown color), and DARBO (in green color) in terms of the number of circuit evaluations. Results for different shot numbers from $\text{shots}=200$ to $\text{shots}=5000$ are reported, respectively. (d) The final converged approximation ratio $1-r$ after sufficient numbers of optimization iterations. 
    For each circuit evaluation, we collect the number of shot measurements to further reconstruct the loss expectation value. Each line is averaged over five w3R graph instances where the shaded range shows the standard deviation of the results across different graph instances. For each graph instance, the best optimization result from 20 independent optimization trials is kept. The error bar in (d) shows the standard deviation across different graph instances. }
    \label{fig:fig3}
\end{figure*}

\subsection{Analytically exact simulation}

\noindent We firstly report the results on optimization trajectories and final converged objective values for analytically exact quantum simulation, which computes the objective function by directly evaluating the expectations from the wavefunction. Fig.~\ref{fig:fig2} shows the results for a collection of different optimizers, including Adam \cite{kingma2015}, COBYLA \cite{Powell1994}, and DARBO. The results are collected from five different $n=16$ w3R graph instances, and for each graph and each optimization method, the best record over $20$ independent optimization trials is reported. The results show the superior efficiency of DARBO over other common optimizers. For all different $p$ values, the approximation gaps $1-r$ of Adam and COBYLA are $1.02\sim 2.08$ times and $1.28\sim3.47$ times larger than those of DARBO, respectively. 
In addition to the efficiency and accuracy, the DARBO performance is also more stable in terms of different problem graph instances. The optimization settings,  five graph instances and the results of more choices of common optimizers are shown in Supplementary Note 1, Supplementary Note 2, and Supplementary Note 3, respectively. We also report the fidelity of three chosen optimizers on different sizes of the graphs in the Supplementary Note 3, which essentially gives the probability of obtaining the exact solution state from measuring the QAOA circuit.

It is worth noting that stability is of great importance for optimizing over the QAOA parameter landscape, as the great number of local minima requires generically exponential independent optimization trials to reach a global minimum \cite{zhou2020quantum}. 
As a result, with the increasing depth $p$ of QAOA, the optimization problem becomes harder, and we see that Adam and COBYLA do not even exhibit a monotonic growth of the accuracy as $p$ increases, despite the fact that the cut size given by the optimal $\vert\psi(\gamma,\beta)\rangle$ (over the choice of $2p$ parameters $(\gamma,\beta)$) is clearly non-decreasing with $p$. The reason of the performance drop is due to a lack of training stability and the sensitivity on initial parameter choices for conventional local optimization methods. DARBO, on the contrary, does give better results with large depth $p$. In other words, one important optimization advantage brought by DARBO is its capability of finding the near optimal parameters with a small number of independent optimization trials, which are not enough to locate the global optimal parameters for conventional optimizers. Some advanced optimization methods such as FOURIER heuristics reported in \cite{zhou2020quantum} can be good at locating global minimum but depend on progressive optimization 
on lower-depth QAOA, leading to a much larger total amount of required circuit evaluations.

\begin{figure*}[htp]
    \centering    \includegraphics[width=1.0\textwidth]{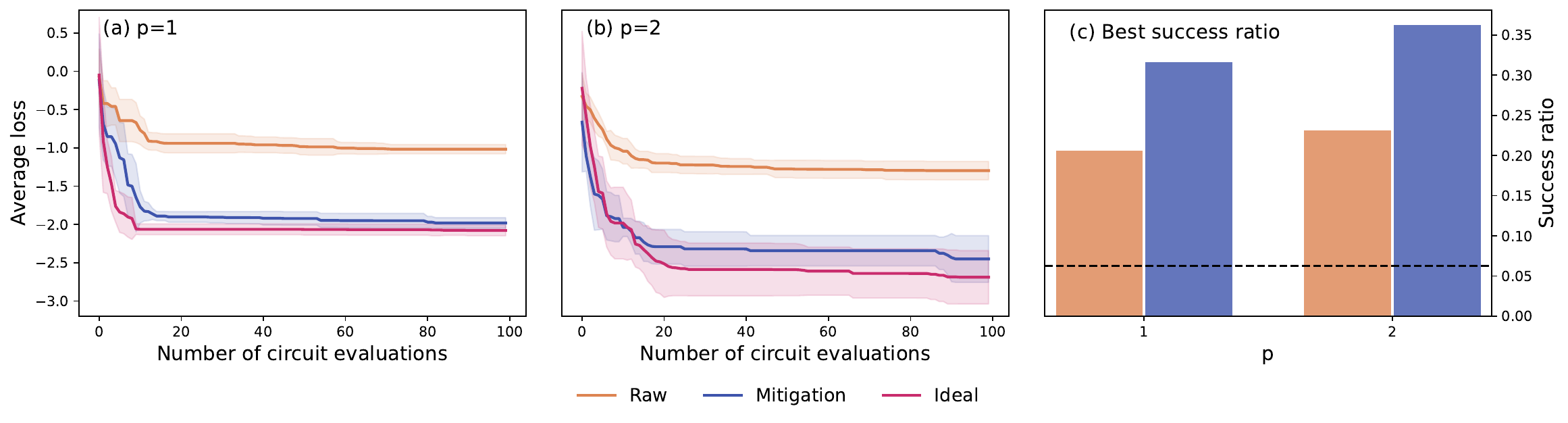}
    \caption{Quantum optimization of a five-variable QUBO problem on real quantum hardware. $\text{Measurement shots}=10000$. (a)-(b) show results from two circuit depths $p=1$ and $p=2$ QAOA, respectively. The line is the average optimization trajectory of five independent optimization trials, while the shaded area represents the standard deviation across five independent optimization trials. Averaged loss refers to the expectation value of the problem QUBO Hamiltonian. Raw (in orange color): at each step, we obtain the loss expectation directly from measurement results on the real quantum processor. Mitigation (in blue color): at each step, we obtain the loss expectation from measurement results integrated with QEM techniques. Ideal (in red color): at each step, we obtain the loss from numerical simulation. (c) The success ratio when we run inference on the trained QAOA program, i.e., the probability that we can obtain a correct bitstring answer for the problem on real quantum hardware. The dashed line is the random guess baseline with a probability of $1/16$. We report the best success ratio of the five optimization trials.}
    \label{fig:fig4}
\end{figure*}

\hspace*{\fill}

\subsection{Numerical simulation with finite measurement shots}
\noindent With the introduction of noise, DARBO shows more advantageous results compared to other optimizers, including Adam \cite{kingma2015}, COBYLA \cite{Powell1994} and SPSA \cite{spall1998}. In Fig.~\ref{fig:fig3}, we show the optimization results of QAOA on five different $n=16$ w3R graph instances with $p=10$ with different measurement shots at each iteration, and for each graph and each optimization method, the best record over $20$ independent optimization processes is reported. For results on QAOA of different $p$, see the Supplementary Note 4. Since QAOA for MAX-CUT has a commutable objective function, the budgets of the measurement shots are all taken on the computational basis. It is natural that with more measurement shots, better accuracy could be achieved for the objective evaluation. Taking $m$ as the given number of measurement shots, we evaluate the final QUBO objective by reconstructing from measurement bitstrings represented by binary valued $z_{ij}=\pm 1$ where $i\leq m$ runs over different shots and $j\leq n$ runs over different qubits. The objective value $C$ is estimated as
\begin{equation}
    C = \frac{1}{m}\sum_{i=1}^m \sum_{j=1}^n \sum_{k=1}^nw_{jk}z_{ij}z_{ik}.
\end{equation}
This value is a random variable with a Gaussian distribution whose mean value is determined by the analytical expectation value of the objective function, and the standard deviation is controlled by the number of total shots $m$ with $\frac{1}{\sqrt{m}}$ scaling.

For DARBO, we can reach a satisfying optimized objective value with a small number of circuit evaluations, e.g., $200$ measurement shots at each round are more than enough. The optimization efficiency gets further improved with an increased number of measurement shots. In Fig.~\ref{fig:fig3}, the approximation gaps $1-r$ of Adam, COBYLA, and SPSA are $4.20\sim 4.59$, $3.95\sim 4.10$, and $4.07\sim 4.79$ times larger than those of DARBO, respectively. Besides, the solution quality from DARBO across different problem instances is impressively stable. The efficiency, accuracy, and stability of DARBO are all much better than those of other optimizers evaluated in our experiments. In addition, the performance gap between DARBO and other optimizers is getting larger compared to the infinite measurement shots (analytical exact) case, which reflects the noise robustness and adaptiveness of our proposed optimizer.

As a gradient-free optimization approach, Bayesian optimization has the advantages of robustness to noise and rigorous reasoning about uncertainty. Na\"ive BO methods have already been utilized to optimize variational quantum algorithms \cite{tibaldi2022bayesian, self2021variational, tamiya2022stochastic, shaffer2023}, but they often suffer from the low efficiency. 
Benefiting from the adaptive trust region and local Gaussian process (GP) surrogate model, DARBO has better potential to optimize noisy problems \cite{eriksson2019scalable}. Instead of directly using the currently best-observed solution $x^*$, we use the observation with the smallest posterior mean under the surrogate model, and therefore the noise affecting $x^*$ has limited effects on the optimization. DARBO is specifically suitable for optimization with shot noises induced by statistical uncertainty of finite measurement shots, since its GP surrogate model assumes that the observations are Gaussian-distributed random variables \cite{gelbart2014bayesian} which is consistent with the case for measurement results with finite shot noises.

\hspace*{\fill}

\subsection{Experiments on superconducting quantum hardware}
\noindent Finally, we run QAOA equipped with DARBO on real quantum hardware to demonstrate its
performance. Quantum error, in addition to shot noise, has a huge impact on optimization performance on real quantum hardware. It has been studied that quantum noises would in general flatten the objective function landscape and induce barren plateaus in variational quantum algorithms
\cite{wang2021noise}.  
Here we investigate the effect of quantum noise on the performance of DARBO for QAOA, and at the same time analyze how the common error mitigation strategies \cite{bravyi2021mitigating,nation2021scalable,temme2017error,li2017efficient} can help in the DARBO process and achieve better end-to-end results. 

The target problem is to optimize a five-variable QUBO (see the Supplementary Note 5 for the problem definition in detail). The experimental results are shown in Fig.~\ref{fig:fig4}.
We carry out the optimization on 1) raw objective value directly evaluated from measurement results on real hardware, 2) mitigation objective value evaluated from measurement results on real hardware with integrated quantum error mitigation techniques including layout benchmarking, readout error mitigation, and zero-noise extrapolation, see Method for more details, and 3) numerical exact value without quantum noise as a comparison. The optimization results are shown in terms of objective optimization history and success ratio from sampling the final QAOA circuit. Although the expectation value is conveniently taken as the objective value for the optimization process,  the success ratio is another important representative metric to straightforwardly evaluate the final performance of QAOA for the QUBO objective since the true objective value can be directly reconstructed by the bitstring measured.

We noticed that DARBO conducted even on raw measurements can improve the cut estimation from the initial value, although it is not good enough compared to the ideal one due to the large influence of quantum noise. The optimization results combined with QEM are much better than the raw evaluations, both in terms of objective evaluation and success ratio from sampling the final QAOA circuit. Moreover, the performance gap between optimization on the mitigation value from experiments and that on the ideal value from numerics becomes larger for larger $p$, which is consistent with the fact that deeper circuits bring larger quantum noise. Still, we show that a deeper QAOA with $p=2$ gives a better approximation of the QUBO objective than a shallow QAOA with $p=1$, achieving a better trade-off between expressiveness and the accumulated noise.

The raw data collected directly from real quantum hardware contain both quantum noises and measure shot noises, which are essentially the bias and variances from the perspective of machine learning. In this QEM-QAOA + DARBO framework, QEM helps to reduce the effects of bias on the hardware (gate noises, readout noises, decoherence noises and so on) by mitigating these errors, and DARBO avoids the negative influence of variances from repetitive measurements (shot noises). Therefore, these two key components together make the proposed framework a powerful optimization protocol for combinatorial optimization problems.

\section{Discussion}\label{sec:discussion}

With a better exploration of the QAOA landscape, the optimization routine based on Bayesian optimization shows weak initial parameter dependence and a better probability of escaping the local minimum. Although, in this work, the dimension of the parameter spaces is still relatively low, an interesting future direction is to generalize similar BO methods from the QAOA setup to other variational quantum algorithms, which has a larger number of parameters. Recently, several advanced BO variants have been proposed to increase the optimization efficiency and robustness in high-dimensional problems \cite{eriksson2021high, letham2020re, nayebi2019framework} and in problems with noisy observations \cite{Letham2017ConstrainedBO, martinez2018practical, frohlich2020noisy, daulton2022robust}. These approaches show superior optimization performances in challenging benchmarks with large parameter sizes and the presence of noises. For instance, an advanced BO approach could efficiently optimize a higher dimensional problem ($D=385$) \cite{eriksson2021high}, and accurately find the best experimental settings for the real-world problems in chemistry \cite{dave2022autonomous}, material sciences \cite{zhang2020bayesian}, and biology \cite{cheng2022odbo}. These cases are potentially relevant for optimization in variational quantum eigensolver, quantum machine learning, and quantum architecture search scenarios. 

We also note that the double-adaptive region idea in BO is a general framework. The detail settings in the DARBO approach could be designed differently for different optimization problems. As a future direction, DARBO algorithm could be extended to include more than two adaptive search regions, and the ranges of these regions themselves could also be adapted during the optimizations.

To successfully scale the QAOA program on real quantum hardware with meaningful accuracy, more pruning and compiling techniques for QAOA deployment, as well as more error mitigation techniques, can be utilized in future works. For example, by differentiable quantum architecture search-based compiling \cite{Zhang2020b}, we can greatly reduce the number of two-qubit quantum gates required with even better approximation performance. There is also QAOA tailored error mitigation algorithm \cite{Weidinger} that trades qubit space for accuracy.

In summary, we proposed an optimizer - DARBO suitable for exploring the variational quantum algorithm landscapes and applied it to the QAOA framework for solving combinatorial problems. The end-to-end performance for combinatorial problems is greatly improved in both numerical simulation and experiments on quantum processors. These promising results imply a potential quantum advantage in the future when scaling up the QAOA on quantum hardware, and give a constructive and generic method to better exploit this advantage.

\section{Methods}\label{sec:methods}
\noindent{\bf Double adaptive-region Bayesian optimization}

In this work, the QAOA problem is formulated as a maximization problem with an objective function of $-\mathcal{L}$ with $D$ total number of parameters to be optimized:
\begin{equation}
    \text{max}_{x \in \mathcal{X}} -\mathcal{L}(\gamma,\beta), 
\end{equation}

Generally, initialized with one randomly selected point from [0,1]$^D$, a Bayesian optimization (BO) algorithm optimizes a hidden objective function $y=y(x)$ over a search space $\mathcal{X}$ by sequentially requesting $y(x)$ on points $x\in \mathcal X$, usually with a single point in each iteration \cite{frazier2018tutorial, mockus2012bayesian}. At each iteration $i$, a Bayesian statistical surrogate model $s$ regressing the objective function is constructed using all currently available data $(x_1, y_1), \ldots, (x_{i-1}, y_{i-1})$. The next point $x_i$ to be observed is determined by optimizing a chosen acquisition function, which balances exploitation and exploration and quantifies the utility associated with sampling each $x \in \mathcal{X}$. This newly requested data $(x_i,y_i)$ is then updated into the available dataset. This BO procedure continues until the predetermined maximum number of iterations (1000 in this work) is reached or the convergence criteria are satisfied. The general BO approach is available in the ODBO package \cite{cheng2022odbo}.

\begin{itemize}
    \item \textit{Gaussian process surrogate model}  
\end{itemize} 

In this study, we use the Gaussian process (GP) \cite{rasmussen2006} as the surrogate model \cite{letham2020re,nayebi2019framework}. With a given set of available observations $(\mathbf{X}, \mathbf{y})$, GP provides a prediction for each point $x'$ as a Gaussian distributed $y' \sim \mathcal{N}(\mu(x'), \sigma^2(x'))$. The predictive mean $\mu(x')$ and the corresponding uncertainty $\sigma(x')$ are expressed as:
\begin{align}
    \mu(x') &= k(x', \mathbf{X})^T \mathbf{K}^{-1} \mathbf{y} \label{eq:mu}\\
    \sigma^2(x') &= k(x', x') - k(x', \mathbf{X})^T \mathbf{K}^{-1} k(x', \mathbf{X}) \label{eq:sigma},
\end{align}
where $k$ is a Mat\'ern5/2 kernel function (Eq.~\ref{eq:matern52}) with a parameter set $\theta=\{\sigma_v, l\}$  to be optimized, and $\mathbf{K} = k(\mathbf{X}, \mathbf{X}) + \sigma_n^2\mathbf{I}$ with a white noise term $\sigma_n$ \cite{rasmussen2006}. In this study, the variance parameter $\sigma_v$ and isotropic lengthscales $l$ constructed by automatic relevance determination in the kernel $k$ are optimized by Adam \cite{kingma2015} implemented in GPyTorch \cite{gardner2018gpytorch}.
\begin{equation}
    k(x, x') = \sigma_v^2(1 + \sqrt{5}r + \frac{5}{3}r^2)\text{exp}(-\sqrt{5}r),
    \label{eq:matern52}
\end{equation}
where $r=\|x-x'\|_2/l$.

The DARBO algorithm is inspired by one of the most efficient BO algorithms, trust region Bayesian optimization algorithm (TuRBO) \cite{eriksson2019scalable}, which performs global optimization by conducting BO locally to avoid exploring highly uncertain regions in the search space. TuRBO was developed to mainly resolve the issues of high-dimensionality and heterogeneity of the problem and has been demonstrated to obtain remarkable accuracy on a range of datasets \cite{eriksson2019scalable}. We have applied TuRBO to QAOA problem and identified its performance advantages. DARBO inherits the advantages of TuRBO with an additional abstraction of the adaptive search region; therefore, it further enhances the optimization efficiency of QAOA problems.

\begin{itemize}
    \item \textit{Adaptive trust region}  
\end{itemize} 

In DARBO, instead of directly querying the next best point for the quantum circuit, we first determine the two adaptive regions, starting from the adaptive trust region. At optimization iteration $i$, the adaptive trust region (TR) is a hyper-rectangle centered with the $i$-th base side length $L_{\text{min}} \leq L_i \leq L_{\text{max}}$ at the current best solution $x^*$ \cite{eriksson2019scalable}. In our case, the minimum allowed length $L_{\text{min}}$ = 2$^{-10}$, and the maximum allowed length $L_{\text{max}}$ = 3.2. To obtain a robust and accurate surrogate for more efficient acquisitions, the GP surrogate model is regressed locally within the trust region, s.t. points far away from the current best solution cannot affect the regression quality. If the most recent queried point is better than the current best solution, we count the query in this iteration as a success. Otherwise, we count it as a failure.
To guarantee that it is small enough to ensure the accuracy of the local surrogate model and big enough to include the actual best solution, the TR (trust region) length $L_i$ is automatically updated with the proceeding of BO cycles as follows:
\begin{align}
    L_0 &= 1.6, \\
    L_i &= \begin{cases}
    \text{min}(L_{\text{max}}, 2L_{i-1}), & \text{if } t_s \geq \tau_s  \\
    L_{i-1}/2, & \text{if } t_f \geq \tau_f
    \end{cases} \ \  (i \geq 1) 
\end{align}
where $\tau_s$ and $\tau_f$ are the threshold hyperparameters for the number of the maximum consecutive successes and that of the maximum consecutive failures, respectively, and $t_s$ and $t_f$ are the actual numbers of consecutive successes and failures in the current BO procedure. We set $\tau_s=3$ and $\tau_f=10$ in this study. If $L_i$ reaches the minimum allowed $L_{\text{min}}$ before the end of the execution, we rescale $L_i$ as $L_i = L_i\times16$. 
The introduction of TR could not only enjoy the traditional benefits of robustness to noisy observations and rigorous uncertainty estimations in BO, but also allow for heterogeneous modeling of the objective function without suffering from over-exploration.

\begin{itemize}
    \item \textit{Adaptive search region}  
\end{itemize} 

We also maintain a second adaptive region, the adaptive search region, with the proceeding of the optimization. The region is automatically determined by the switch counter $c_s$, which counts the consecutive searching failure number in the current search region. Once $c_s$ reaches the maximum allowed consecutive failure hyperparameter $\kappa_s$ = 4, the adaptive search region switches to the other predetermined searching region. This also indicates that exploitation within this current region might be currently exhausted. The adaptive search regions serve as constraints for the parameters $x$. Only the points in the current searching region will be considered as possible candidates to be queried, and the switch counter allows BO search with different constraints.
Inspired by the conclusion from \cite{zhou2020quantum} that the parameter space can be reduced in given graph ensembles
the two adaptive search regions are determined as $A=$[$-\pi/2$, $\pi/2$]$^D$ (the restricted search space) and $B=$[$-\pi$, $\pi$]$^D$ (the full search space) in our study. 

Note that the two adaptive regions take different roles in the DARBO algorithm. The adaptive trust region provides a more precise surrogate model around the best solution by limiting the training points to be fitted in GP, while the adaptive search region constrains the candidate parameter sets temporarily by switching between the restricted search space and the full search space. In this work, to search more efficiently, we further restrict the acquisition function to select new candidate points that lie in the overlap between the TR and the adaptive search region, as in the default implementation of the ODBO package \cite{cheng2022odbo}. For the cases where there is no overlap between the adaptive trust region and the adaptive search region, we reset the trust region to be equal to the current adaptive search region. 

\begin{itemize}
    \item \textit{Upper confidence bound acquisition function}  
\end{itemize}

In order to query the next best point, acquisition functions that balance exploitation and exploration using the posterior distributions from GP (Eq. \eqref{eq:mu} and \eqref{eq:sigma}) are required. The point with the highest acquisition value is the candidate point to be queried from the quantum circuit. In this study, we only evaluate the points within the adaptive search region using upper confidence bound (UCB) \cite{srinivas2009gaussian} acquisition function in Eq. \ref{eq:ucb}.
\begin{equation}
    \alpha_{UCB}(x) = \mu(x) + \beta\sigma(x), \label{eq:ucb}
\end{equation}
where $\beta = 0.2$ is a predefined hyperparameter to control the degree of exploration, and $\mu$ and $\sigma$ are the predictive mean and uncertainty from local GP modeled with points in the adaptive trust region.

\hfill{}

\noindent{\bf Quantum Error Mitigation}

Besides the quantum algorithm, another key to operating experiments on quantum devices is the investigation and mitigation of quantum errors. We utilize a number of error mitigation methods in order to obtain desirable results for our QAOA program.

\begin{itemize}
    \item \textit{Layout benchmarking}  
\end{itemize} 

The qubit quality and the single- and two-qubit gate fidelity vary across different quantum devices and vary over time. Device error can be initially attenuated by selecting qubits with better quality and links that host two-qubit gates with a lower error rate. These metrics can be benchmarked and collected by calibration experiments, including $T1 /T2$ characterization and randomized benchmarking. In particular, we chose two-qubit gates that are directly connected on hardware to avoid additional swap manipulations introduced in quantum compiling.

In order to further determine the circuit structures, especially the applying order of those two-qubit couplings (all two-qubit couplings commute with each other in QAOA for QUBO objectives), we run multiple reference circuits by permuting those two-qubit gates under the same set of parameters and identify the optimal circuit structure that shows the highest fidelity with the ideal state. These trial experiments provide valuable insights on the circuit structures with overall low noise effects that balance the influence of crosstalk and circuit depth. The key tradeoff in layout benchmark is that: on the one hand, for compact two-qubit gate layout, the overall circuit depth is short, while there are more two-qubit gates applied at the same time which may induce larger cross-talk effect. On the other hand, the two-qubit layout can be placed in a rather sparse fashion, which has less cross-talk effects but takes longer physical evolution time. Therefore, we can explore different two-qubit layouts to minimize the overall noise effect. In our implementation, we use brute-force search. For systems with larger sizes, greedy search or more advanced reinforcement learning methods can be explored for better scalability, which is an interesting future direction.

\begin{itemize}
    \item \textit{Readout error mitigation}  
\end{itemize} 

The imperfect measurement operation on a quantum circuit can result in readout errors that bias the original quantum state to certain bit strings. The readout error on the device used in the experiments is around $10^{-1}$. We mitigate the readout error by several steps: 1) learn how the readout is biased by measuring states that produce fixed bitstring outputs, 2) encode all deviations in a confusion matrix, and 3) invert the confusion matrix and apply it to raw counts of bitstrings to correct the measurement bias. The size of the confusion matrix is $2^n$ where $n$ is the number of measured qubits. For the error learning process, we tried both local learning and global learning modes. The local learning process characterizes the readout bias on each single qubit independently (involving $2$ calibration circuits in the minimal case), while the global learning  process models the readout bias of the Hilbert space expanded on all the qubits (involving $2^n$ calibration circuits) by capturing the readout correlation between qubits. We find that the local learning is good enough in our experiments as the readout correlation is negligible on the device we used.

\begin{itemize}
    \item \textit{Zero-noise extrapolation}  
\end{itemize} 

Zero-noise extrapolation (ZNE) is one of the most widely used error mitigation methods that can be applied without detailed knowledge of the underlying noise model and exhibits significant improvement in the results evaluated on quantum devices.  The main idea of ZNE is to obtain expectation values at several different error rates and extrapolate to the zero noise limit according to those noisy expectation values. Suppose that two-qubit gates contribute the most of the errors, we conduct experiments on different error rates $[1,3,5]$ by locally folding those two-qubit gates $[U, UU^\dagger U, UU^\dagger UU^\dagger U]$ to avoid circuit depth that challenges the coherence time. As for the experiments in the main text, for $p=1$ ($p=2$), we adopt linear (quadratic polynomial) extrapolation to estimate the mitigated results. All the expectation values used in ZNE are firstly mitigated by readout error mitigation.

All the numerical simulations and quantum hardware experiments including error mitigation in this work are implemented and managed using TensorCircuit \cite{Zhang2022}---a high-performance and full-featured quantum software framework for the NISQ era.

\section*{Data availability}
All graphs and results presented in this study are shared on a github repository (\url{https://github.com/sherrylixuecheng/EMQAOA-DARBO}). The additional figures for test results are shown in the Supplementary Information and the full statistics of the optimized losses and $r$ values are included in a separate excel (Supplementary Data 1)

\section*{Code availability}
The entire DARBO framework is available on github (\url{https://github.com/sherrylixuecheng/EMQAOA-DARBO}) with example jupyter notebooks and all the testing results. The example codes to perform QAOA evaluations and DARBO optimizations are also available on TensorCircuit (\url{https://github.com/tencent-quantum-lab/tensorcircuit})\cite{Zhang2022}, and ODBO (\url{https://github.com/tencent-quantum-lab/ODBO})\cite{cheng2022odbo}.

\section*{Competing interests}
The Authors declare no Competing Financial or Non-Financial Interests.


\section*{Author contributions}
L.C., Y.C., and S.-X.Z. developed the idea, implemented the simulation and experiments, analyzed the results, and wrote the manuscript. L.C. performed the numerical experiments. L.C., Y.C.,  S.-X.Z. and S.Z. participated in the discussion for the research and revising the manuscript.

\section*{Acknowledgements}
We appreciate Jonathan Allcock's helpful discussions and  Zhaofeng Ye's suggestions on the schematic workflow's graphics design. We also thank the technical support from Maochun Dai, Zhenxing Zhang, and Dengfeng Li for the usage of the quantum device.

\end{document}